\documentclass[11pt,onecolumn]{article}% \usepackage{multicol}
\usepackage{graphics} \usepackage[final]{graphicx}
\usepackage{graphicx}

\usepackage{graphicx}
\usepackage{amsfonts, amssymb}
\usepackage{latexsym}
\usepackage{amsmath}

\newtheorem{theorem}{Theorem}[section]

\newtheorem{definition}[theorem]{Definition}

\newtheorem{proposition}[theorem]{Proposition}
\newtheorem{example}[theorem]{Example}

\newcommand{\bi}{\begin{itemize}}
\newcommand{\ei}{\end{itemize}}
\newcommand{\ben}{\begin{enumerate}}
\newcommand{\een}{\end{enumerate}}
\newcommand{\beq}{\begin{equation}}
\newcommand{\eeq}{\end{equation}}
\newcommand{\beqa}{\begin{eqnarray}}
\newcommand{\eeqa}{\end{eqnarray}}

\usepackage[tight,footnotesize]{subfigure}
\begin{document}

\title{Geometric WOM codes and coding strategies for multilevel flash memories
\thanks{This work was supported in part by the National Security Agency under Grant Number H98230-11-1-0156. The United States Government is authorized to reproduce and distribute reprints not-withstanding any copywrite notation herein.}}

\author{   
Kathryn Haymaker and Christine A. Kelley\\
Department of Mathematics\\ 
University of Nebraska-Lincoln \\
Lincoln, NE 68588\\ 
}

%\author{Kathryn Haymaker and Christine A. Kelley }
%
%\authorrunning{Haymaker and Kelley}  % if too long for running head

% the affiliations are given next; 
% select one of the mail options.
% don't give your e-mail address
%\email{s-khaymak1@math.unl.edu, ckelley2@math.unl.edu }
%\email{fauthor $|$ sauthor@example.com}
%\email{}

%\institute{Department of Mathematics\\ University of Nebraska-Lincoln\\ Lincoln, NE 68588, USA.}

\maketitle

\linespread{1.0}
\selectfont

\begin{abstract}

This paper investigates the design and application of write-once memory (WOM) codes for flash memory storage. Using ideas from Merkx \cite{m84}, we present a construction of WOM codes based on finite Euclidean geometries over $\mathbb{F}_2$. This construction yields WOM codes with new parameters and provides insight into the criterion that incidence structures should satisfy to give rise to good codes. We also analyze methods of adapting binary WOM codes for use on multilevel flash cells. In particular, we give two strategies based on different rewrite objectives. A brief discussion of the average-write performance of these strategies, as well as concatenation methods for WOM codes is also provided. 
\end{abstract}


%\vspace{-0.2in}

%%%%%%%%%%%%%%%%%%%%%%%%%%%%%%%%%%%%%%%%%

\section{Introduction}
\label{intro}

Non-volatile flash memories are becoming increasingly popular due to their 
potential for high throughputs and low power consumption. Flash memory 
storage is a technology that is based on organizing the memory into blocks 
of cells in which each cell can be charged up to one of $q$ levels. While 
increasing the charge of a cell is easy, decreasing the charge is costly since the entire block containing the cell must be erased and rewritten.  Such an operation involves reprogramming roughly $10^5$ cells. Moreover, frequent 
block erasures also reduce the lifetime of the flash device. 
 It is therefore desirable to be able to write as many times as 
possible before having to erase a block \cite{rs82, fs84, kysvw10, csbb07}.   Like any storage 
device, the flash cells are also prone to errors due to charge leakage or the writing process.  Thus, the coding design goals for flash memories include 
maximizing the number of writes between block erasures, correcting cell 
charge leakage errors, and correcting errors that occur during the writing process.

An information theoretic approach to writing on memories with defects was first considered by Kuznetsov and Tsybakov \cite{kt74}, and later surveyed in \cite{kh94}. The write-once memory (WOM) model, introduced by Rivest and Shamir \cite{rs82}, and other constrained memory models (WUM, WIM, WEM) can be considered as particular cases of the general defective channel \cite{kh94, az89, c88}. Due to the asymmetric costs associated with increasing and decreasing cell levels, the flash memory model can be viewed as a generalization of the WOM model. 
As a result, WOM codes have been proposed for flash cells having two levels (i.e., capable of storing one bit of information per 
cell) \cite{j07, jb08, yvsw08}. Error-correcting codes for the general defective channel and for WOM have also been considered, although addressing errors while incorporating rewriting capabilities is difficult, and  
many codes in the literature are optimized primarily for one of these goals 
\cite{csbb07, jlw09, jb09, jb08, zc91, j07}. 

This paper is organized as follows. The rest of this section provides notation and background on codes for flash memories. In Section 2, after summarizing  Merkx's construction of WOM codes from finite projective geometries,  we present a 
 new  construction of
WOM codes using finite Euclidean geometries. In Section 3 we explore methods of adapting binary WOM codes for multilevel flash cells, and introduce two strategies that achieve this with respect to different goals.  We also examine the average write analysis of these strategies for two specific WOM codes. Finally, we summarize ways to combine WOM codes with classical error correcting codes using concatenation in Section 4. We conclude the paper in Section 5 with some future directions.

%%%%%%%%%%%%%%%%%%%%%%%%%%%%%%%%

%\section{Preliminaries}
%\label{ prelim}

%\vspace{-0.05in}
\subsection{Preliminaries}

We now give some definitions and notation that will be used in this paper. A write-once memory (WOM) is a storage device over a binary alphabet where a zero can be increased to a one, but a one cannot be changed back to a zero. An information message is encoded and stored in a string of cells in the memory, referred to as a \emph{ cell state vector}\footnote{This terminology was introduced in \cite{ j07} in reference to the structure of flash memory, but it is convenient to use in the WOM case as well.}. The cells in the cell state vector form the codeword and can be updated, or \emph{rewritten}, to represent a different message. Only the most recently written message is retained.  

A write-once memory code is composed of a set $V$ of information words, called \emph{variable vectors}, and a set $S$ of cell state vectors with $S\subseteq \mathbb{F}_2^n$, corresponding to the codewords of the WOM code. Many different cell state vectors can represent the same information message. In addition, the WOM code is equipped with an encoding and decoding function. The encoding function takes as inputs both the current state of the memory and the new information message to be stored. Specifically, it maps the current cell state vector to an updated cell state vector that represents the new information message and is component-wise greater than or equal to the previous state. The decoding function maps the resulting cell state vector to the updated information message. The amount of information messages that can be encoded at each time step need not be the same, however, as the following notation conveys. 
\begin{definition}
Let $\langle v_1, \ldots, v_t\rangle/n$ denote a $t$-write WOM code on $n$ cells, where $v_i$ is the number of messages that can be represented on the $i^{th}$ write. 
In the fixed information case, i.e., when $v_1=\cdots = v_t$, such a WOM code will be denoted by $\langle v\rangle^t/n$. \end{definition}

The \emph{rate} of a WOM code is  \[ R=\frac{ \log_2(v_1\cdots v_t)}{n}.\]

When $q = 2$, the flash cell is called a 
single level cell (SLC) since the  cell can only represent one nonzero value, and a multilevel cell (MLC) when $q > 2$ 
 as it can store values $\{0,1,2,\ldots, q-1\}$. Note that an SLC can store one bit of information per cell whereas an MLC can store multiple bits of information per cell. Fiat and Shamir considered a generalized version of a WOM, in which the storage cells have more than two states with transitions given by a directed acyclic graph \cite{fs84}. The idea of extending to multilevel cells was further explored by Jiang in \cite{j07}, in which he considered generalizing error-correcting WOM codes. Techniques for rewriting codes on $q$-ary cells include floating codes, which were introduced by Jiang, Bohossian, and Bruck \cite{jbb07}, and more generally, trajectory codes, which are described in \cite{jlsb09}. Although these are similar objects, we will use the term {\em flash codes}, introduced in \cite{yvsw08}, to refer to a rewriting code on multilevel cells. 
  
\begin{definition}
When $q>2$,  %$W(n,M,t)_q$ 
$\langle v\rangle^t_q/n$ will denote a $t$-write \emph{flash code} for use on 
cells having $q$ levels, where the code has block length $n$ and 
$v$ messages can be represented at each write. The capacity of a flash memory is the maximum number of writes possible for $n$ number of cells, $v$ number of information messages to be represented in each write, and $q$ number of levels per cell. 
\end{definition}

Fu and Han Vinck \cite{fh99} showed that the maximum total number of information bits that can be stored per cell over $t$ writes is at most
\[ \log_2(1+(q-1)t). \]

The next example, from \cite{rs82}, gives the canonical example of a WOM code. 
\begin{example} The Rivest and Shamir WOM code is shown in Table \ref{r-s-WOM} \cite{rs82}. It maps two information bits to three 
coded bits and is capable of tolerating two writes. 
Note that any of the four messages may be written at 
either write. The table is interpreted as follows: on the first write, the encoding function takes the current all-zero state and the new information message and maps it to the representation of that message in the `first write' column. On the second write, the encoding function takes the current cell state and the new information message and outputs the cell state vector opposite the new message in the `second write' column. For example, the message sequence $01 \rightarrow 11$ would be recorded as $100 \rightarrow 110$. If the new information message is the same as the information represented by the current cell state vector, the memory remains unchanged. Decoding is as follows: the cell state vector $(a_1, a_2, a_3)$ can be decoded as $((a_2+a_3) \mbox{ (mod 2) }, (a_1+a_3) \mbox{ (mod 2)})$. 

{\scriptsize
\begin{table}
\label{r-s-WOM}
\begin{center}
\begin{tabular}{c|c|c}
\text{Information} \ &\ $1^{st}$  write  \ & \ $2^{nd}$  write \\
\hline
%\vspace{.1in} 
00&000&111\\
01&100&011\\
10&010&101\\
11&001&110\\
\end{tabular}
\vspace{.2 in}
\caption{$\langle 4 \rangle^2/3$ WOM-code by Rivest and Shamir.} 
\end{center}
\end{table} }

\label{rivest-shamir-WOM}
$\hfill \Box$
\end{example}

\section{Finite Geometry WOM-codes}

In this section, we apply ideas from \cite{m84} to design WOM codes based on finite Euclidean geometries. 
We first provide some relevant definitions. 

\begin{definition}
The \emph{$m$-dimensional Euclidean geometry over $\mathbb{F}_2$}, denoted by $EG(m,2)$, is an incidence structure with $2^m$ points and $2^{(m-1)}(2^{m}-1)$ lines. The points in $EG(m,2)$ may be regarded as all $m$-tuples over $\mathbb{F}_2$, and each pair of points defines a line.  

\end{definition}

Note that the set of points in $EG(m,2)$ forms an $m$-dimensional vector space over $\mathbb{F}_2$. A $\mu$-flat in $EG(m,2)$ is a $\mu$-dimensional subspace of the finite geometry, defined next. We will use the term {\em hyperplane} to refer to a subspace of dimension $m-1$ in either $PG(m,2)$ or $EG(m,2)$. 

\begin{definition} Let $X$ be the set of points in $EG(m,2)$.  
A \emph{ $\mu$-flat} in $EG(m,2)$ passing through a point $a_0$ consists of points of the form $a_0+ \beta_1 a_1+\cdots + \beta_{\mu}a_{\mu}$, where $a_0, \ldots, a_{\mu}\in X$ are linearly independent and $\beta_1, \ldots, \beta_{\mu}\in \mathbb{F}_2$. 
\end{definition}

The number of $\mu$-flats in $EG(m,2)$ is
\[  2^{(m-\mu)} \prod_{i=1}^{\mu} \frac{2^{(m-i+1)}-1}{2^{(\mu-i+1)}-1}. \] 
Moreover, each $\mu$-flat in $EG(m,2)$ is a coset of an $EG(\mu, 2)$, and thus contains $2^{\mu}$ points. 

\begin{definition}
The \emph{finite projective geometry of dimension $m$ over $\mathbb{F}_2$}, denoted $PG(m,2)$, is an incidence structure with $2^{m+1}-1$ points and $\frac{(2^{m+1}-1)(2^m-1)}{3}$ lines. The points are the nonzero $(m+1)$-tuples $(a_0, a_1, \ldots, a_m)\in \mathbb{F}_2^{m+1}$, and a line through two distinct points $a_0$ and $a_1$ contains exactly the set of points $\{a_0, a_1, a_0+a_1\}$.   
\end{definition} 

For more details, see \cite{lc04} and \cite{ms79}.

  Merkx constructed a family of WOM codes based on the $m$-dimensional finite projective geometries over $\mathbb{F}_2$ \cite{m84}. The construction exploits a connection between the binary Hamming codes and $PG(m,2)$ that allows the WOM codes to be decoded via syndrome decoding. Specifically, the minimum weight codewords of the $[2^{m+1}-1, 2^{m+1}-m, 3]$ Hamming code $\mathcal{C}$ generate $\mathcal{C}$ and correspond to the incidence vectors of lines in $PG(m,2)$. In Merkx's construction, the messages correspond to points in the geometry. The WOM codewords, i.e. the cell state vectors, are a subset of $\mathbb{F}_2^{m+1}\setminus \mathcal{C}$, and thus, since the Hamming code is perfect, these codewords are always one error from a binary Hamming codeword. The location of the error indicates the point in the geometry that corresponds to the information message.  
  
  \begin{figure}
\centering{\resizebox{3.8in}{1.3in}{\includegraphics{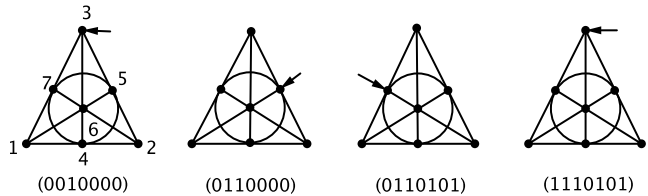}}
\vspace{-.2in}
\caption{Four writes using the Merkx $PG(2,2)$ WOM code.}}
\label{fig:Fano_Example}
\end{figure} 

  \begin{example}
  The $PG(2,2)$ WOM code of \cite{m84} is a $\langle 7\rangle^4/7$ code. Each position of a codeword corresponds to a point of the Fano Plane, and each codeword is the incidence vector of a substructure of the geometry that highlights a particular point being represented. 
  Codewords are incidences of the following: on the first write, a point on the Fano Plane; on the second write, a line missing a point; on the third write, a line with a point off of it; on the final write, either the union of two lines or the plane missing a point. Thus to decode the WOM code, Merkx observed that syndrome decoding identifies the information message. Figure 1 shows the write sequence $3\rightarrow 5 \rightarrow 7 \rightarrow 3$ using the $\langle 7 \rangle^4/7$ code from the Fano Plane. The arrow indicates the information point and the corresponding cell state vector representing that information is listed below each write. Note that the sequence of cell state vectors is monotonically increasing in each component as the writes progress.
  \label{PG2-2-WOM}
   
$\hfill \Box$

\end{example}

The following proposition, by Cohen, Godlewski, and Merkx in \cite{cgm86}, formulates more precisely the parameters of the WOM codes that result from this construction method. 

\begin{proposition}
The number of writes that can be attained with a length $2^m-1$-WOM code, storing $m$ bits on each write, is $2^{m-2}+2$. 
\end{proposition}

\subsection{WOM codes from $EG(m,2)$} 
We now extend Merkx's idea and design WOM codes from $EG(m,2)$. 
Since Hamming codes are punctured Reed-Muller codes, and are given by geometric designs over the binary field, a construction similar to the method above can be applied to $EG(m,2)$. Minimum weight codewords also generate the $r^{th}$ order Reed-Muller code $\mathcal{R}(r,m)$, of length $2^m$, and correspond to $(m-r)$-flats in the Euclidean geometry $EG(m,m-r)$. Analogous to the Merkx construction, we will use the connection between minimum weight words in $\mathcal{R}(m-2, m)$ and the planes in $EG(m,2)$ to construct our WOM code. The codewords are designed to be Hamming distance one away from a codeword of $\mathcal{R}(m-2, m)$, and thus are incidence vectors of configurations of points in the Euclidean geometry. Such substructures include a point, a plane with a point missing, and a plane with a point off of it. These WOM codes may be decoded using any Reed-Muller decoding technique. 

The next two examples illustrate this construction for $m=3$ and $m=4$. 

\begin{example}
\label{EG3-2-WOM}
Using $EG(3,2)$, the resulting code is an $\langle 8,8,8,4\rangle/8$ WOM code. In other words, the code attains four writes on eight cells, where eight possible messages can be stored in the first three writes, and four messages can be stored in the fourth write. Recall that $EG(3,2)$ has eight points, $28$ lines, and $56$ planes. Each message corresponds to one of the points in the geometry. On the first write, a message $i \in \{1,\ldots, 8\}$ is represented by 
a weight one cell state vector, where the one is in the $i^{th}$ coordinate. On the second write, a weight three cell state vector indicates a plane with a point missing, where the missing point is the information message. On the third write, the ones in the cell state vector correspond to a plane with a point off of it, where the point off the plane is the message. Observe that on each of the first three writes, it is possible to represent any of the eight messages. Finally, on the fourth write, only messages corresponding to positions of the cell state vector with entry zero can be represented (except for the message represented in the third write, which can always remain on the fourth write, if needed). If $i$ is one of these messages, then to represent $i$ on the fourth write, the cell state vector will have a one in every coordinate except position $i$. 

 As an example, the message sequence $1\rightarrow 3 \rightarrow 2\rightarrow 7$ is demonstrated in Figure \ref{eg32writes}. 

\begin{figure}
\centering \tiny
\subfigure[Write 1.]{\resizebox{1.1in}{1.1in}{\includegraphics{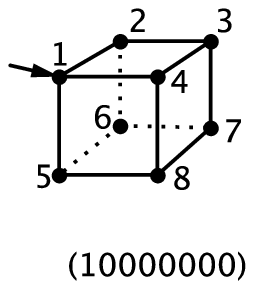}}}
\subfigure[Write 2.]{\resizebox{1.1in}{1.1in}{\includegraphics{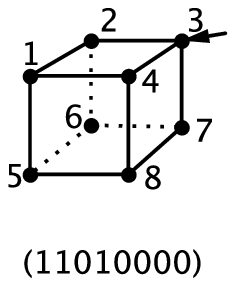}}}
\subfigure[Write 3.]{\resizebox{1.1in}{1.1in}{\includegraphics{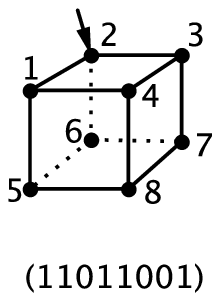}}}
\subfigure[Write 4.]{\resizebox{1.1in}{1.1in}{\includegraphics{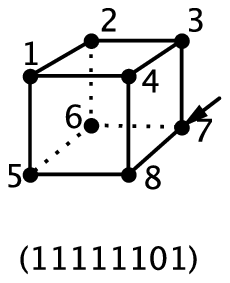}}}
\caption{The message sequence $1\rightarrow 3 \rightarrow 2\rightarrow 7$ in the $EG(3,2)$ WOM code.}
\label{eg32writes}
\end{figure} 
 
 $\hfill \Box$

 \end{example}

 In constructing the WOM code from $EG(3,2)$, it is not possible to represent more than four messages on the fourth write. Indeed, after the third write, the cell state vector contains five ones and three zeros, so at most $ \log_2(3) $ information bits can be conveyed by the remaining zero-valued positions. The message that is stored in the third write can always be represented on the fouth write, simply by leaving the memory state unchanged. Thus, one of at most four messages can be represented on the fourth write.

 \begin{example}
 Using $EG(4,2)$, the resulting WOM code has parameters  \[ \langle 16, 16, 16, 12, 8, 8, 8, 4 \rangle /  \ 16. \] 
 Recall that $EG(4,2)$, shown in Figure 3, has 16 points and 140 planes, and can be partitioned into two parallel $3$-flats. The first four writes are the same as in Example \ref{EG3-2-WOM}, by using the $EG(3,2)$ code on a $3$-flat that contains the points corresponding to the first  four information messages. After the fourth write, the points in that $3$-flat are all programmed to one, and the $EG(3,2)$ WOM code may be applied to the points of the remaining $3$-flat to encode the final four writes. 
 
% \vspace{-.1in}
$\hfill \Box$
 \end{example}

 \begin{figure}
\centering{\resizebox{1.5in}{1.5in}{\includegraphics{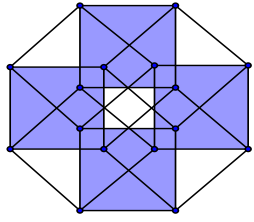}}
\vspace{-.2in}
\caption{$EG(4,2)$, with four parallel planes shaded, as in \cite{ms79}.}}
\label{eg42}
\end{figure} 
%\vspace{-0.15in}

\begin{proposition}
 The $EG(m,2)$ WOM code achieves $4(m-2)$ writes and has parameters

\[ \langle \underbrace{ \strut 2^m, 2^m, 2^m, 2^m - 4,  2^{m-1}, 2^{m-1}, 2^{m-1},  2^{m-1} -4, \ldots,  8, 8, 8,   4 }_{4(m-2)} \rangle /   2^m. \]

\end{proposition} 

{\em Proof:} 
The cell state vector has length $2^m$, equal to the number of points in $EG(m,2)$. Recall that each cell state vector in the $EG(m,2)$ WOM code will be  Hamming distance one away from a codeword of the Reed-Muller code $\mathcal{R}(m-2, m)$. We proceed by induction on the dimension of the finite geometry. The base case is the $EG(3,2)$ WOM code. Now suppose that there exists an $EG(k,2)$ WOM code with the parameters described in Example \ref{eg32writes}. Consider the finite Euclidean geometry $EG(k+1, 2)$. Note that $EG(k+1, 2)$ can be partitioned into two parallel hyperplanes, i.e. two disjoint copies of $EG(k,2)$. Since any four points lie on a common hyperplane (in fact, many), there exists a hyperplane that contains the points that correspond to the first four information messages to be written. These messages can be encoded using the $EG(3,2)$ WOM code on a cube within this hyperplane containing those points. After the first four writes, all points in the hyperplane are set to one, and the $EG(k,2)$ code can be used on the remaining hyperplane. Thus, this $EG(k+1, 2)$ WOM code allows for $4((k+1)-2)$ writes, and has the parameters listed above, with $m=k+1$. 
$\hfill \Box$

Since codewords of the WOM code are Hamming distance one from a codeword of the corresponding Reed-Muller code, performing syndrome decoding on a stored cell state vector will provide the location of the position of the ``error". The code is designed so that this position corresponds to an information message, i.e., a point in the geometry. Thus, syndrome decoding identifies the message, and can be used to decode the $EG(m,2)$ WOM code.

\subsection{Comparison}

Table \ref{geom-WOM-table} shows the rates of the proposed $EG(m,2)$ WOM codes and the $PG(m,2)$ WOM codes from \cite{m84} for small values of $m$. As expected from the geometric structure, the efficiency of the $EG$ WOM codes is less than that of the $PG$ codes. Indeed, when $m = 2$ and $3$, the $PG(m,2)$ WOM codes have been shown to be optimal \cite{cgm86}. However, the construction presented here yields a new family of WOM codes that have simple encoding and decoding algorithms,  and shows that  variable information WOM codes may also be obtained from incidence structures.  

In general, designing efficient WOM codes from incidence structures requires low weight incidence vectors, and intersections of these structures that can point to specific messages. In the case of $EG(m,2)$, the $(m-2)^{th}$ order Reed-Muller code was chosen so that the corresponding minimum weight codewords  would be planes and therefore have low weight.  Since any two distinct planes intersect in $0$ or exactly $2$ points, taking unions of multiple planes does not uniquely designate any one particular point when multiplicity is considered. The authors are interested in using other structures that may be exploited in designing WOM codes where multiplicity can be incorporated, and are currently working on designing WOM codes from general bipartite graphs using insights gained from the the rewriting rules of the geometric constructions.

{\scriptsize
\begin{table}
\begin{center}
\begin{tabular}{c|c|c}
\mbox{ \ Code \ }&\mbox{ length }&\mbox{ rate } \\
\hline
PG(2,2)&7& 1.60\\
  EG(3,2)&8&1.38 \\
PG(3,2)&15& 1.82\\
 EG(4,2)&16& 1.66\\
PG(4,2)&31& 1.60\\
 EG(5,2)&32& 1.50 \\
\end{tabular}
\vspace{.2 in}
\caption{Comparison of rates of small dimension projective and Euclidean geometry WOM codes.}
\label{geom-WOM-table}
\end{center}
\end{table}}

\section{Using binary WOM codes on multilevel cells}

The development of flash memory cells on $q>2$ levels has renewed interest in efficient coding strategies for `generalized' write-once memories, i.e., those with greater than two states per cell. Applying binary WOM codes for use on multilevel cells provides a basis for comparison for efficient multilevel coding schemes.   
In this section we examine construction methods for adapting binary 
WOM codes for use on multilevel cells.

{\scriptsize
\begin{table}
\begin{center}
\begin{tabular}{c|c|c|c|c}
$x$&$\mu^{1}(x)$&$\mu^{2}(x)$&$\mu^{3}(x)$&$\mu^{4}(x)$\\
\hline
00&000&111&111&222\\
01&100&011&211&122\\
10&010&101&121&212\\
11&001&110&112&221\\
\end{tabular}
\vspace{.2 in}
\caption{Rivest-Shamir code adapted to $q=3$ levels.}
\label{rivest-shamir-WOM-table}
\end{center}
\end{table}
}

One way to use binary codes\footnote{
The idea of reducing the cell state vectors modulo 2 was 
also used in \cite{hla11} to adapt {\em classical} codes for use on multilevel cells.} on 
$q$-level cells is to read the cells modulo 2. One naive approach is to let the set of codewords consist of all cell-state vectors that reduce modulo 2 to a binary codeword. 
A more efficient application of a $\langle v\rangle^t/n$ code to $q$-level 
cells is to increase the charge of all cells to $1$ after the $t^{th}$ 
write, and then employ the code again. We will refer to this scheme as the {\em complement scheme}, since reduction modulo 2 either reveals a WOM codeword or the complement of a codeword. More precisely, in the complement scheme, let $x$ denote the information message, and $c^i(x)$ be a codeword that represents $x$ on the $i^{th}$ write. We reuse the binary WOM code by taking 
$c^{t+i}(x)=c^i(x)+{\bf 1}$, for $i<t$, where 
${\bf 1}$ is the all ones vector. Similarly, after $mt$ writes,  
the cell values are increased to $m$, and we set $c^{mt+k}(x)=c^{k}(x)+m\cdot{\bf 1}$ for 
$k=1, \ldots, t-1$. Note that this scheme guarantees $(q-1)t$ writes. Table \ref{rivest-shamir-WOM-table} shows Example \ref{rivest-shamir-WOM} adapted to $q=3$-level 
cells in this way.

We will use this simple scheme as a basis for comparison when considering the following methods of adapting binary WOM codes to $q$-levels. \\

\noindent {\bf Construction:}
Consider a $\langle 2^k\rangle^t/n$ WOM code.  Let $x$ be a binary information 
sequence of length $k$, and let $U(x)=\{u: u=c^{i}(x) \text{ for some } i=1,\ldots, t\}$. Let $s$ be a length $n$ cell 
state vector representing the message $x$. Given $s$, suppose we want to write a 
new message $y \ne x$. Let $V$ be the set of $n$-tuples with all entries even 
(possibly 0) and less than $q$. We present two strategies.
 
\begin{itemize}
\item {\bf Strategy A}: To minimize the number of cells that are 
increased, search the set $U(y)+V$ for the 
representation whose difference from $s$ requires the fewest cells to 
increase. 
Thus, look for $s'\in U(y)+V$ such that 
$s'\ge s$ (componentwise, all entries in $s'$ are at least as much as those in 
$s$) and further that $s$ and $s'$ differ in the least number of places, i.e. 
the Hamming weight, $wt_H(s'-s)$ is minimized. The new cell state vector is $s'$ and 
represents the new message $y$. In searching the set $U(y)+V$ as the cell values approach $q$, we omit the values of $s'$ that would cause a block erasure.

\item {\bf Strategy B}: To minimize the magnitude of the resulting cell 
state vector $s'$, search the set $U(y)+V$ for 
the representation whose difference from $s$ is such that the maximum cell entry 
of $s'$ is minimized. If there is a tie, arbitrarily choose one that requires the 
fewest number of cells to increase. 
Thus, look for $s'\in U(y)+V$ such that 
$s'\ge s$ and that the maximum entry in $s'$ is the smallest. 

\end{itemize}
 
\label{ WOM_new}

For specific codes, the strategies can be described more explicitly. For example, the following flash code encoding map is based on Example \ref{rivest-shamir-WOM},  and uses reduction modulo 2 to 
identify the decoding map from the cell state vectors to the variable vectors.  
Following Strategy A, the rewriting rule is as follows. Let $s$ be the current 
cell state vector representing the message $x$, and $y$ the new message to be written.
\begin{itemize}
\item If $x,y \in \mathbb{F}_2^2 \setminus \{00\}$, 
\begin{itemize}
\item If $s$ mod  2 $= c^{1}(x)$,  add the weight one vector 
$w=c^{2}(y)-c^{1}(x)$ to 
the current state, to obtain the new cell state vector $s'$ = 
$s+w$.
\item If $s $ mod   2$ = c^{2}(x)$
write $w=c^{1}(z)$, where $z\in 
\mathbb{F}_2^2\setminus\{00, x, y\}$, to obtain $s'=s+w$.
\end{itemize}
\item If $x = 00$, write $c^{1}(y)$.
\item If $y = 00$, then if $s$ mod  2 $= c^{1}(x)$, add $c^{1}(x)$ to $s$; otherwise add ${\bf 1} - c^{2}(x)$ to $s$.
\end{itemize}

Following Strategy B, the rewriting rule depends on the actual magnitude 
(in $\{0,\ldots, q-1\}$) of each cell entry. 

The 
general rule is to increase a subset of the cells such that the new vector 
reduces to either $c^{1}(y)$ or $c^{2}(y)$ modulo 2 and no one cell is allowed 
to gain too much charge. 

\begin{example} Using the rules above for the Rivest-Shamir WOM code in Example \ref{rivest-shamir-WOM}, suppose the following information sequence is to be 
stored in a given set of cells with $q = 4$ levels. \[ 11 \rightarrow 00 
\rightarrow 01  \rightarrow 10 \rightarrow 11 \rightarrow 
01\]

%\vspace{-0.1in}

Following Strategy A, the sequence of cell state vectors is as follows 
\[A: 001 \rightarrow 002 \rightarrow 102 \rightarrow 103 \rightarrow 
203 \rightarrow 213 \]

%\vspace{-0.1in}

Following Strategy B, the sequence of cell state vectors is as follows 
\[B: 001 \rightarrow 111 \rightarrow 211 \rightarrow 212 \rightarrow 
312 \rightarrow 322 \]

%\vspace{-0.1in}

$\hfill \Box$ 

\end{example} 

\begin{example}
To further illustrate the different strategies, consider writing the sequence $1\rightarrow2\rightarrow 1 \rightarrow 3$ using the $PG(2,2)$ WOM code in Example \ref{PG2-2-WOM}, where the labeling on the Fano Plane is as in Figure 1. 
Following Strategies A and B, the sequence of cell state vectors is as follows:

\[A: (1000000) \rightarrow (1001000) \rightarrow (1002000) \rightarrow (1002001)  \]

\[B: (1000000) \rightarrow (1001000) \rightarrow (1001101) \rightarrow (1101111) \]

$\hfill \Box$ 
\end{example}

\subsection{Analysis of Strategies A and B}

The expected number of writes for floating codes was studied in 
\cite{flm08, cflm10}
and can be more important than the worst case analysis in determining which codes to use in practice. 
Code constructions in \cite{jbb07} have a guarantee of $(q-1)+\lfloor \frac{q-1}{2}\rfloor$ writes for a $k=2$-dimensional message space and $n=2$ cells. The same paper also proved the existence of floating codes that achieve $(q-1)n-o(n)$ writes as $n\rightarrow \infty$ for fixed $k$ and $q$. Asymptotically optimal codes for the average case with $k=2$ have been constructed where the expected number of writes grows like $n(q-1)-o(q)$ \cite{cflm10}. Both cases include the assumption that only one cell level changes at each write, which is reasonable when $n\gg 2^k$. However, since Strategies A and B are intended to be used for {\em any} WOM code, not just those that meet this criterion, we do not use this assumption.

The guaranteed number of writes using Strategy B for the $\langle 4\rangle^2/3$ Rivest-Shamir WOM code on $q$ level cells is $2(q-1)$. This can be seen by examining a sequence of messages that cause a maximum number of cell increases under Strategy B. For example, the alternating  sequence of messages 
$00\rightarrow 01\rightarrow 00\rightarrow 01\rightarrow 00 \rightarrow \dots \rightarrow  01\rightarrow 00$ has cell state vector sequence $000 \rightarrow 100\rightarrow 111\rightarrow 211 \rightarrow 222 \rightarrow \dots \rightarrow (q-1)(q-2)(q-2)\rightarrow (q-1)(q-1)(q-1)$. Observe that for every two writes,  the cell state vector does not increase a cell level more than once, and both representations of a given message are used. Thus, the guaranteed number of writes using Strategy B is $2(q-1)$. 

The guaranteed number of writes using Strategy A for the $\langle 4\rangle^2/3$ Rivest-Shamir WOM code on $q$ level cells is also $2(q-1)$. Again we consider a sequence of messages that causes the maximum number of cell increases. For example, the alternating sequence of messages 
$00\rightarrow 01\rightarrow 00\rightarrow 01\rightarrow 00 \rightarrow \dots \rightarrow  01\rightarrow 00 \rightarrow 01$ has cell state vector sequence $000 \rightarrow 100\rightarrow 200\rightarrow 300 \rightarrow 400 \rightarrow \dots \rightarrow (q-2)00\rightarrow (q-1)00 \rightarrow (q-1)11\rightarrow (q-1)22\rightarrow \dots \rightarrow (q-1)(q-1)(q-1)$.  Observe that the first  $q-1$ writes follow the Strategy A protocol to increase the fewest number of cells, but that once any cell attains the maximum charge, the Strategy continues to write using the next best representation choice for each message. Thus, a total of $2(q-1)$ writes are guaranteed.

The following theorem shows that the guaranteed number of writes for both Strategies A and B is at least as good as the complement scheme for any general binary WOM code.

\begin{theorem}
Let $\mathcal{C}$ be a $\langle v\rangle^t/n$ binary WOM code. Then, the guaranteed number of writes by applying either  Strategy A or Strategy B to $\mathcal{C}$ on $q$-level flash cells is at least $(q-1)t$.
\label{guaranteed}
\end{theorem}

{\em Proof:} We prove by induction on $q$. For $q=2$, the WOM code already guarantees $t$ writes. So  assume the hypothesis holds for $q=r$. That is, for any sequence of messages, we are guaranteed at least $(r-1)t$ writes using Strategy A or B. Now let us consider the case when $q=r+1$. Then for any sequence of $(r-1)t$ messages, using Strategy A or Strategy B, by the induction hypothesis we will reach a cell state vector $(c_1,c_2,\dots,c_n)$, with entries $c_i \le r-1$, $i=1,2,\dots,n$. 
We can now artificially increase each cell levels to $r-1$  at the end 
of $(r-1)t$ writes to yield a cell state vector $(r-1,r-1,\dots,r-1)$. Without loss of generality, the 
cell state vector $(r-1, r-1, \cdots,r-1)$ can be thought of as being the 
all-zero vector $(0,0,\cdots,0)$.  It is now easy to see that either 
Strategy A or Strategy B will allow us to write at least $t$ more times  
using the original $t$ writes of the binary WOM code $\mathcal{C}$. Thus, 
a total of $rt$ writes is guaranteed for either Strategy when $q=r+1$, 
thereby proving the result. $\hfill \Box$ \\  \vspace{0.1in}

To see if the lower bound of $(q-1)t$ writes is met in Theorem \ref{guaranteed},  the weight distributions of the different representations for each message in the original WOM code have to be taken into account. For example, for two write WOM codes where the minimal weight representation for each message is unique, the guaranteed number of writes is $2(q-1)$ as above. The authors are currently looking at how to classify when a WOM code meets this lower bound using the weight distributions of the message representations.

\begin{figure}[!h]
\centering{\resizebox{4.0in}{2.5in}{\includegraphics{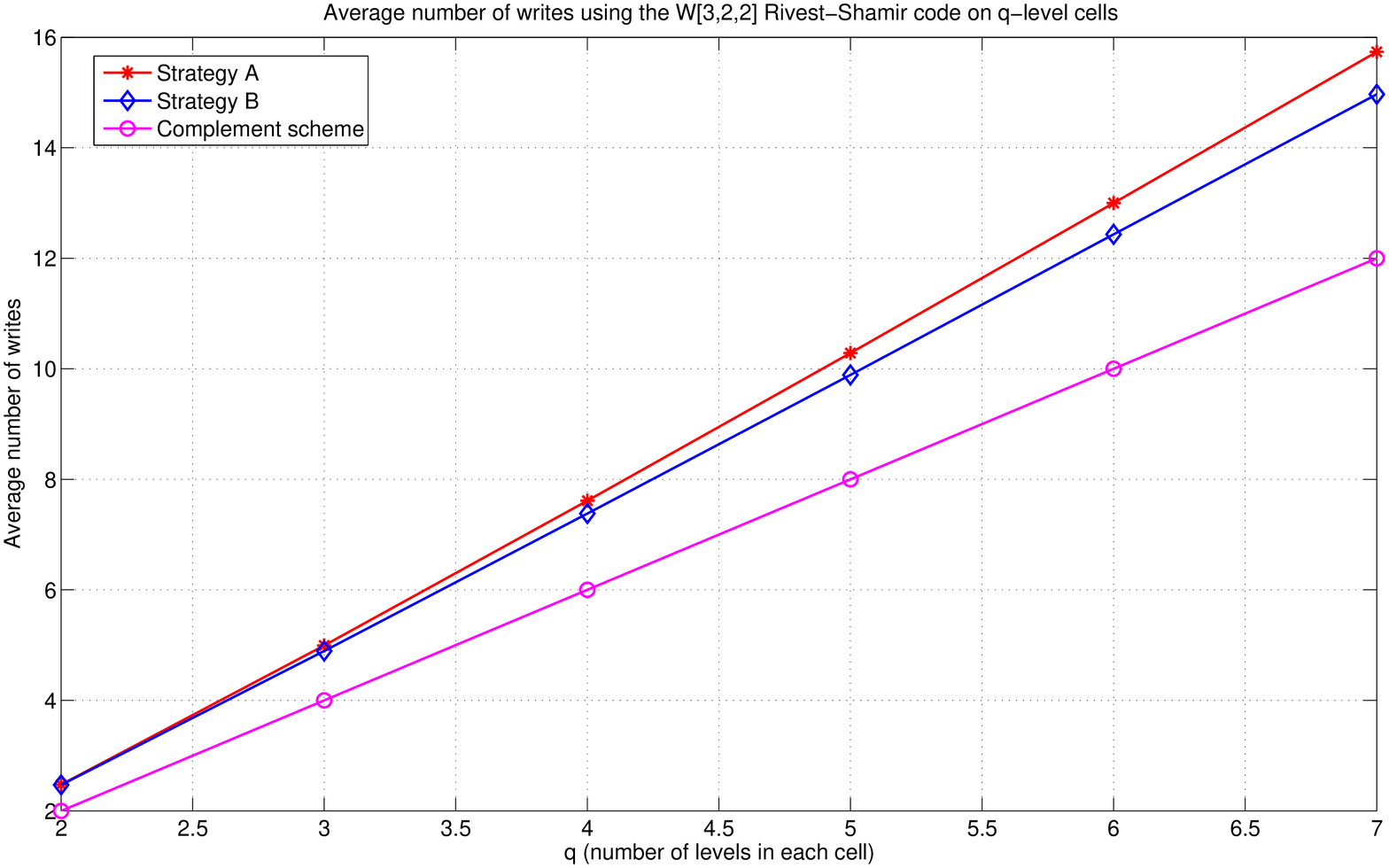}}}
\caption{Comparison of the average number of writes achieved by Strategies A and B and the complement scheme.} 
\label{ave_writes}
 \end{figure}

Strategies A and B applied to the Rivest-Shamir code each guarantee two writes when $q = 2$ and four writes when $q = 3$, whereas the expected number of writes using the Strategies for this code (assuming a uniform distribution on the message space) is approximately 2.47 for $q = 2$ and 4.89 for $q = 3$ for each case. Note that the simple application of the Rivest-Shamir code to $q$-level cells using the complement scheme requires $q \ge 3$ to get more than two guaranteed writes. Figure \ref{ave_writes} compares the average number of writes of the complement scheme, Strategy A, and Strategy B on $q$-level cells when applied to the binary Rivest-Shamir WOM code from Example \ref{rivest-shamir-WOM}. In Monte Carlo simulations, $10^5$ random message sequences were generated and the number of writes was recorded for the three different methods. As shown in Figure \ref{ave_writes}, the strategies applied to the Rivest-Shamir code exhibit a noticeable gain over the the complement scheme that is growing as $q\rightarrow \infty$. However, the average number of writes for each Strategy is still quite far from the capacity limit on the number of writes possible for representing four messages per write using three cells on $q$-levels (see Section 1). 

Strategies A and B did not exhibit much gain over the complement scheme when the $PG(2,2)$ code in Example \ref{PG2-2-WOM} was simulated for small $q$. We believe that this is due to the near-optimality of the $PG(2, 2)$ WOM code. Further, we believe that in general, the more optimal a code is, the less it will benefit from the strategies, since the reapplication of the code under the complement scheme already generates an efficient code. 

 In \cite{cflm10}, two coding schemes are presented that have a similar flavor to Strategies A and B, but apply in the different setting of random floating codes. In that work, the authors propose two random coding schemes: a ``Simple scheme" that randomly chooses to increase a single cell by one, and a ``Least scheme" that chooses a message representation that increases the coordinate with the lowest charge level. In contrast, Strategies A and B in this paper apply to any WOM code without the assumption that only one cell increases at each write. We also expect that the performance of codes under these strategies will differ more for certain classes of codes and when non-uniform distributions on the message space are considered. Further analysis of the performances of the strategies for different WOM codes is underway, including quantifying their average performance using both uniform and nonuniform distributions on the message space.  

%%%%%%%%%%%%%%%%%%%%%%%%%%%%%%%%%%%%%%%%%%%%%%%%%%%%%

\section{Concatenated error-correcting flash codes}

In this section we consider ways that code concatenation may be used to obtain new WOM or flash codes.  
Let $[n,k,d]_q$ denote a classical $q$-ary linear code of block length 
$n$, dimension $k$, and minimum distance $d$. Two classical codes may be concatenated as follows.

\begin{definition} Let $\mathcal{A}$ be an $[n_1,k_1,d_1]_{q^{k_2}}$ code and $\mathcal{B}$ be an $[n_2,k_2,d_2]_q$ code. Then the concatenated code $\mathcal{C} = \mathcal{A}\boxtimes\mathcal{B}$ is an $[n_1n_2,k_1k_2,d_1d_2]_q$ code with outer code $\mathcal{A}$ and inner code $\mathcal{B}$. 
 The $k_1$ information symbols (each chosen from a $q^{k_2}$-ary alphabet) are 
first encoded into $n_1$ symbols using  $\mathcal{A}$. Each of the encoded 
symbols is then represented by $k_2$ $q$-ary symbols. Each group of these $k_2$ 
symbols is then encoded into $n_2$ $q$-ary symbols using  $\mathcal{B}$. Thus, 
$n_1n_2$ encoded symbols are obtained to form a codeword in $\mathcal{C}$. 
\label{ concatenation_classical_classical} \end{definition}

The above concatenation may be seen by the following mapping
\[ \mathbb{F}_{q^{k_2}}^{k_1}\  \  \underrightarrow{\ \ \ \mathcal{A}\ \ \ } \  \  \mathbb{F}_{q^{k_2}}^{n_1}\  \  \underrightarrow{\mbox{$q$-ary representation}} \ \ \mathbb{F}_q^{n_1k_2} \  \  \underrightarrow{\ \ \ \mathcal{B}\ \ \ }\ \  \mathbb{F}_q^{n_1n_2} \]

Concatenating classical codes with binary WOM or flash codes 
 yields codes with both error correction and rewrite 
capabilities.

Several researchers have observed that an outer  $\langle 2^k\rangle^t/n$ WOM code $\mathcal{A}$ when concatenated with an inner 
$[m,1]_{2}$ repetition code  $\mathcal{B}$ yields a $\langle 2^k\rangle^t/nm$ binary WOM code 
$\mathcal{C} = \mathcal{A}\boxtimes\mathcal{B}$, where $C$ can correct $\lfloor\frac{m-1}{2}\rfloor$ errors \cite{zc91, ysvw10, j07}. We expand on these ideas to obtain codes for multilevel flash cells. 

A code $C_W\boxtimes C_R$, where $C_W$ is a WOM code and $C_R$ is a 
length-$m$ repetition code, can be employed as an error-correcting code on 
$q$-level cells with the following strategy: on the first write, the binary 
codeword is written on the cells. An error can be detected by majority decision among each set 
of $m$ consecutive positions. For subsequent writes and error correction, we will 
read the $q$-ary vector as a binary codeword from $C_W$, by reducing the values in 
the cells modulo 2. In particular, if a one was erroneously written on the first 
write in a cell that should have contained a zero, we correct the error by 
increasing the level of the cell to $2$, which is viewed as a 0 (modulo 2). 
The error has been corrected in the binary word that is read, and the code can correct $\lfloor\frac{m-1}{2}\rfloor$ errors on each write. Subsequent 
writes are achieved by increasing chosen cell levels to obtain the desired 
parity, modulo 2. 

The following theorem uses this method 
to obtain an error-correcting WOM code.   Note that errors can occur in either direction and are assumed to be of magnitude one.

\begin{theorem} Let $C_W$ be a $\langle 2^k\rangle^t/n$ WOM code and let $C_R$ be 
the $[m,1,m]_2$ repetition code. The code $C_W \boxtimes C_R$ 
is an $\langle 2^k \rangle^t/mn$ $\lfloor\frac{m-1}{2}\rfloor$-error-correcting WOM-code 
on SLCs. Moreover, applied to $q$-level cells and using the 
reduced binary vector representation,
 ${C_W} \boxtimes C_R$ is a $\langle 2^k \rangle_q^{t'}/mn$  
 flash code, where $t'=\lceil \frac{(q-1)t}{3}\rceil$ and $\lfloor\frac{m-1}{2}\rfloor$ errors can be corrected at each write. 
\label{basic_thm}
\end{theorem}

{\em Proof:} 
For $q = 2$ the resulting code is a $\langle 2^k \rangle^t/mn$ $\lfloor \frac{m-1}{2}\rfloor$-error correcting WOM code. For any $q$, the length $mn$-code has 
dimension $k$. We show that the worst-case number of rewrites is 
$\lceil \frac{(q-1)t}{3}\rceil$. Note that $C_W\boxtimes C_R$ is still a binary code, 
but we use it on the $q$-ary cells by reading the information stored in 
the cells via the reduced binary vectors. Up to $\lfloor\frac{m-1}{2}\rfloor$ errors can 
be detected and corrected at each write. Note that in this scheme, error 
correction consists of increasing the charge level of the cell by one to 
correct the parity in that entry of the reduced binary vector. In the worst 
case, an error occurs in the same position on every write, and so that position sees an increase of three levels at each write. However, in the absence of errors we could 
achieve $(q-1)t$ writes due to the rewriting capability of $C_W$ and the 
reapplication of the WOM code on $q$-level cells. Thus, the worst-case number of writes in 
the error case is $\lceil \frac{(q-1)t}{3}\rceil$. $\hfill \Box$

As an example of the reading process, if $q=4, n=1, m=3$, the 
sequence $(332)$ in a cell-state vector would be read as $(110)$ in 
$C_W\boxtimes C_R$, and decoded to $(111)$ using majority rule. As an example of the error-correction process, consider a cell that is meant to be increased to $0$ (modulo 2); if an error causes the cell to instead be read as $1$ (modulo 2), then to correct it the charge is increased again. Thus that cell has seen a total increase of three levels on that write cycle. 
A similar idea of increasing the cell levels to correct for errors has also been considered in \cite{j07, jlw09}. 

\begin{example}  
Let $C_W$ be the $\langle 4 \rangle^2/3$ WOM code defined in Example 
\ref{rivest-shamir-WOM} and let $C_R$ be the $[3,1,3]_2$  repetition 
code. Then the code $C_W \boxtimes C_R$ is a $\langle 4 \rangle^2/9$  
single error-correcting WOM code on SLCs (first observed in \cite{zc91}). Moreover, on 
$q$-level cells, the code $C_W \boxtimes C_R$ is a $\langle 4\rangle_q^{\lceil\frac{2(q-1)}{3}\rceil}/9$ single error-correcting flash code. $\hfill \Box$
\end{example}

\begin{example} Let $C_W$ be the $\langle 7\rangle^4/7$ code based on 
$PG(2,2)$ from  \cite{m84} and  let $C_R$ be the $[3,1,3]_2$ binary 
repetition code. Then the code $C_W \boxtimes C_R$ is a $\langle 7 \rangle^4/21$  
single error-correcting WOM code on SLCs. 
Moreover, on $q$-level cells, the code $C_W \boxtimes C_R$ is a $\langle 7 \rangle_q^{\lceil \frac{4(q-1)}{3}\rceil}/21$ single error-correcting flash code. 
$\hfill \Box$
\end{example}

We next show how to obtain a flash code with increased 
error-correction by concatenating an inner flash code with an 
outer classical code.

\begin{theorem} Let $C_1$ be an $[n_1, k_1]_{q^{k_2}}$ code that corrects $e$ 
errors, and $C_2$ a $\langle 2^{k_2}\rangle_q^t/n_2$ $E$-error-correcting WOM code. 
Then $C_1\boxtimes C_2$ is a $\langle 2^{k_1k_2}\rangle^t_q/(n_1n_2)$ WOM code capable of 
correcting ${\bf (E+1)(e+1) - 1}$ errors.
\label{ classical_WOM_thm}
\end{theorem}

{\em Proof:} The length and dimension of $C_1\boxtimes C_2$ is immediate.
Note that this code 
achieves $t$ writes since the inner flash code is capable of $t$ writes. 
The minimum number of errors that must occur for a decoding failure is 
$(E+1)(e+1)$, where $E+1$ errors occur among each of
$e+1$ distinct length-$k_2$ $q$-ary expansions of symbols in $C_1$. Any 
smaller number of errors can be corrected by the length $n_1n_2$ 
concatenated code. $\hfill \Box$

For comparison, we show the concatenation of a inner binary repetition  
code with a classical binary outer code for use on $q$-level flash cells.

\begin{theorem}
Let $C$ be an $[n,k,d]_2$ $e$-error-correcting code and let $C_R$ be the 
$[2m+1,1,2m+1]_2$ binary repetition code. Then the code $C \boxtimes C_R$ 
for $q$-level cells results in a $\langle 2^k \rangle^t_q/((2m+1)n)$ flash code that corrects $(me+m+e)$ errors and guarantees $t=\lceil \frac{q-1}{3} 
\rceil$ writes. 
\label{classical_classical_thm}
\end{theorem}

{\em Proof:}
The length and dimension follow from the construction. Concatenating two binary codes results in a binary code, but we use reduction modulo 2 to adapt the code to $q$-ary cells. Errors that result in a change in parity of a cell can be corrected by increasing the level of the cell by one. In the worst case, an error occurs in the same cell at every write. In order to correct it, the cell level is increased by one so that it has the same parity as the entry before the error occurred. Thus this code guarantees $\lceil \frac{q-1}{3} \rceil$ writes. Note that the outer code can correct up to $e$ errors and the inner code can correct up to $m$ errors. Thus, the concatenated code can tolerate $(m+1)(e+1)-1=me+m+e$ errors.   
$\hfill \Box$

Observe that this use of a classical code on multilevel cells gives better 
error-correction capabilities than the code in Theorem \ref{basic_thm} but 
can tolerate fewer rewrites since the only rewrite capabilites come from 
the number of levels.

\begin{example} Let $C$ be an $[n,k,d]_2$ $e$-error-correcting code and 
let $C_R$ be the $[3,1,3]$ binary repetition code. Then the code $C \boxtimes 
C_R$ for $q$-level cells yields a $\langle 2^k\rangle^t_q/(3n)$ flash code that corrects ${\bf 2e+1}$ errors and gets $t=\lceil \frac{q-1}{3} 
\rceil$ writes. 
$\hfill \Box$
\end{example}

%%%%%%%%%%%%%%%%%%%%%%%%%%%%%%%%%%%%%

\section{Conclusions}

We showed how the structure of finite Euclidean geometries can be used to obtain new variable information WOM codes. We also introduced several strategies for adapting WOM codes to multilevel cells that allow for a 
greater rewrite capability than classical codes adapted for multilevel coding 
schemes. 
Combined with concatenation, these codes also have the ability to correct 
multiple errors. We are currently investigating the use of other incidence structures, including finite geometries over $\mathbb{F}_q$, for developing new coding schemes for multilevel flash memories.

\subsection*{Acknowledgments}
Some of this work was completed while the authors were at the Fall 2011 program on {\em Combinatorial, Algebraic and Algorithmic Aspects of Coding Theory} at the Bernoulli Center, EPFL, Switzerland. The authors thank the organizers and administrative staff for their hospitality, and the anonymous reviewers for their helpful comments.

\end{document}